# Collective spin waves in arrays of Permalloy nanowires with single-side periodically modulated width

G. Gubbiotti,[1] L.L. Xiong,[2] F. Montoncello[3] and A.O. Adeyeye[2]

[1]Istituto Officina dei Materiali del CNR (CNR-IOM), Sede Secondaria di Perugia, c/o Dipartimento di Fisica e Geologia, Università di Perugia, I-06123 Perugia, Italy

[2]Information Storage Materials Laboratory, Department of Electrical and Computer Engineering, National University of Singapore, 117576 Singapore.

[3] Dipartimento di Fisica e Scienze della Terra, Università di Ferrara, Via G. Saragat 1, I-44122 Ferrara, Italy

ABSTRACT

We have experimentally and numerically investigated the dispersion of collective spin waves propagating through arrays of longitudinally magnetized nanowires with periodically modulated width. Two nanowire arrays with single-side modulation and different periodicity of modulation were studied and compared to the nanowires with homogeneous width. The spin-wave dispersion, measured up to the third Brillouin zone of the reciprocal space, revealed the presence of two dispersive modes for the width-modulated NWs, whose amplitude of magnonic band depends on the modulation periodicity, and a set of nondispersive modes at higher frequency. These findings are different from those observed in homogeneous width NWs where only the lowest mode exhibits sizeable dispersion. The measured spin-wave dispersion has been satisfactorily reproduced by means of dynamical matrix method. Results presented in this work are important in view of the possible realization of frequency tunable magnonic device.

The experimental evidence that dense arrays of parallel ferromagnetic nanowires (NWs) magnetized along their length and separated by air gaps support the propagation of spin waves (SWs) along the periodicity direction, [1] dates to more than ten years ago. From that time, arrays of longitudinally magnetized NWs represented a model system for investigating the collective SWs in dipolarly coupled magnetic elements. In this configuration, the complexity associated to the inhomogeneity of the internal magnetic field are not present, and the SW spectrum is relatively simple: there are few modes whose frequency oscillates in certain frequency range (magnonic bands) separated by frequency regions where propagation is not allowed (band gaps). [10,11,12] In most of the cases, the lowest frequency mode, only, presents a significant dispersion and can be used for carrying and processing information in magnonic devices. [2]

Later on, several types of NW arrays have been investigated, including NWs with: magnetization-modulation,[3,4] asymmetrically sawtooth-shaped notched,[5] alternated width,[6,7] bend sections,[8,9] localized ion implantation,[10] two alternating ferromagnetic material [11,12] and layered structure where two ferromagnetic layers are either exchange [13,14,15] or dipolarly coupled. [16]

More in detail, micromagnetic and experimental studies on width-modulated NWs were exclusively performed to understand the SW propagation properties along their length and how it is affected by the profile of the width modulation. [17,18,19,20]

Despite the large number of studies, the experimental evidence of effect of modulation on the dispersion of SWs propagating perpendicular to the NW length is still lacking in the literature.

In this letter, we have investigated by Brillouin light scattering spectroscopy the dispersion of collective SWs in dense arrays of $Ni_{80}Fe_{20}$ (Permalloy, Py) single-side width modulated NWs with different modulation periodicity and results are compare with those obtained for homogeneous NWs without any width modulation. The SW dispersion was studied by BLS over three Brillouin zones (BZs) of the reciprocal space by sweeping the wave vector parallel to the direction of the array period (perpendicularly to the NWs length). It has been found that in width-modulated NWs two dispersive modes are detected in the lowest frequency range of the spectrum whose amplitude of oscillation depends on the modulation periodicity.

Periodic Permalloy NW arrays with fixed width $w$=350 nm were fabricated over an area of 4x4 mm$^2$ on top of an oxidized silicon substrate using deep ultraviolet (DUV) lithography at 193-nm exposure wavelength leading to resist NW arrays. [21]The width modulation is along the right edge of the NW with modulation periodicity $p$=1000 nm ($NW_{SM-1000}$) and $p$=500 nm ($NW_{SM-500}$). The intrusion has triangular shape with rounded corners and depth fixed at 90 nm. In the narrowest (widest) NW region, the width is $w$=260 nm (350 nm) and consequently the interwire distance is $d$=210 nm (120 nm). The array periodicity is $a$=$w$+$d$= 470 nm, resulting in the edge of the first BZ located at $\pi/a = 0.66 \times 10^7$ rad × m$^{-1}$. An array of NWs with homogeneous (NW$_{NM}$) width $w$=350 nm and $d$=120nm is also fabricated and used as reference sample. Scanning electron microscope (SEM) images of the homogeneous width NW and of the arrays with different modulation periodicity are shown in Fig. 1. Hysteresis loops, measured by magneto-optic Kerr effect (MOKE) in the longitudinal configuration and with the magnetic field applied along the NW length, have a squared shape with 100% remanence and with increasing coercivity passing from 132 Oe (NW$_{NM}$) to 165 Oe for (NW$_{SM-1000}$) and finally to 242 Oe (NW$_{SM-500}$).

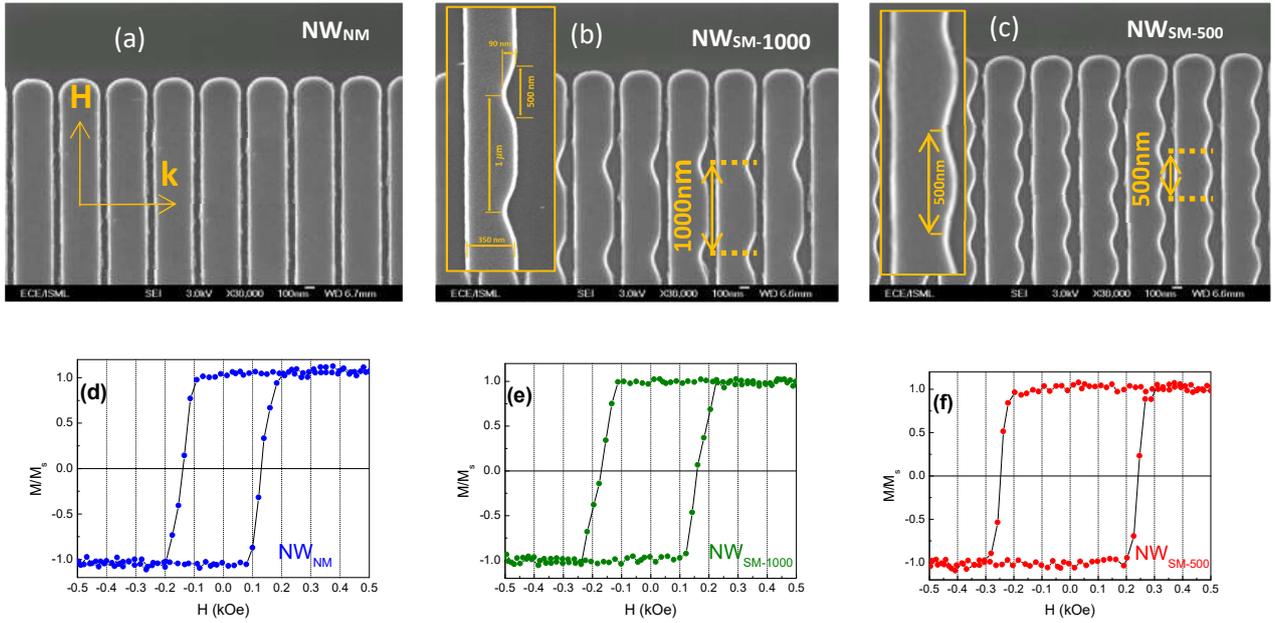

fig. 1 (a) SEM images of the investigated NWs arrays without (NW$_{NM}$) and with (b,c) width modulation with periodicity ($p$) along their length of (b) 1000 nm (NW$_{SM-1000}$) and (c) 500 nm (NW$_{SM-500}$). Inset of panel (a) shows the direction of the applied magnetic field (H) and of the wave vector ($k$) while insets of panels (b) and (c) show a magnification of the width modulated NWs and report the typical dimensions of the notch. Panels (d,e and f) show the measured MOKE loops for the three investigated NW arrays.

A monochromatic laser beam of wavelength λ= 532 nm is focused on the sample surface and the backscattered light is analyzed in frequency by using a (3+3)-tandem Sandercock-type Fabry–Pérot interferometer.[22] An external magnetic field H=500 Oe is applied along the NWs length, which ensures the NW saturation, see Fig. 1, and the wave vector $k$ was swept along the array periodicity direction, perpendicularly to the magnetic field direction in the range between 0 and 2×10$^5$ rad/cm, which corresponds to map the SW dispersion up to the third BZ of the reciprocal space. The $k$-vector magnitude is selected by changing the incidence angle of light (θ) upon the sample according to the relation $k = (4\pi/\lambda) \times \sin(\theta)$. [23]

The dynamical matrix method (DMM),[24] was used to calculate both the frequencies and the spatial profiles of all magnetic modes as a function of the Bloch wavevector. [25,26,27]

The equilibrium magnetization configuration of the NWs has been calculated at H=500 Oe by using the OOMMF micromagnetic code with periodic boundary conditions [28]: for each sample, the primitive periodic cell was discretized by micromagnetic cells of $5 \times 5 \times 30$ nm$^3$. The magnetic parameters were used: saturation magnetization $M_s$ =750 G, gyromagnetic ratio ($\gamma/2\pi$) 2.95 GHz/kOe and exchange stiffness constant $A = 1.0 \times 10^{-6}$ erg/cm. Then, DMM computed the full set of modes at each wavevector value within ΓY direction of the Brillouin zone, with steps of $0.1 \times (2\pi/a)$. The solutions $\delta \boldsymbol{m}$ found by the DMM are written as Bloch waves: $\delta \boldsymbol{m}_k = \delta \widetilde{\boldsymbol{m}}_k e^{i\boldsymbol{k} \cdot \boldsymbol{r}}$, where $\delta \widetilde{\boldsymbol{m}}_k$ is the cell function, which has the periodicity of the array, and $\boldsymbol{k}$ is the wavevector in the Brillouin zone. In the following, we will plot the real z-component of $\delta \widetilde{\boldsymbol{m}}_k$ for $\boldsymbol{k}$=0 (Γ), which is mainly responsible for the BLS cross section. [29]

Fig. 2 presents the measured BLS spectra for the three samples investigated at the center (Γ-point, $k$=0) and the boundary (Y-point, $k=\pi/a$) of the first BZ. At $k$=0, only one mode ($f_1$) is visible for the homogeneous width NWs (NW$_{NM}$) while a doublet of peaks is observed for the width-modulated NWs ($f_1$ and $f_1'$). Remarkably for the NW$_{SM-1000}$ array, these two peaks have comparable intensity while for the NW$_{SM-500}$ the lowest frequency peak is less intense than the higher frequency one. Moreover, the intensity asymmetry between the Stokes and anti-Stokes side of the spectra increases passing from $k$=0 to $k=\pi/a$ as already discussed for both continuous films [30] and patterned structures.[29]

In addition, either at $k$=0 or at $k=\pi/a$, the frequency difference between these two modes ($f_1$-$f_1'$) increases from 1.1 GHz to 2.2 GHz when one considers the NW$_{SM-1000}$ and the NW$_{SM-500}$ array, respectively. This difference is mainly due to the significant variation of the lowest frequency mode ($f_1$) on the $p$ while the frequency ($f_1'$) is almost insensitive on this parameter.

For $k=\pi/a$, other peaks appear in the measured BLS spectra at higher frequency with respect to those observed for $k$=0 because of their non-vanishing cross-section at finite $k$-values. [27] It is noteworthy that the intensity ratio between the $f_1$ and $f_1'$ peaks changes for the NW$_{SM-1000}$ array while for NW$_{SM-500}$ it remains almost constant.

The frequency increase of all the peaks is clearly visible passing from the center to the boundary of the BZ. This significant frequency variation is the fingerprint that collective SWs of Bloch type are propagating through the NWs array. [1,12]

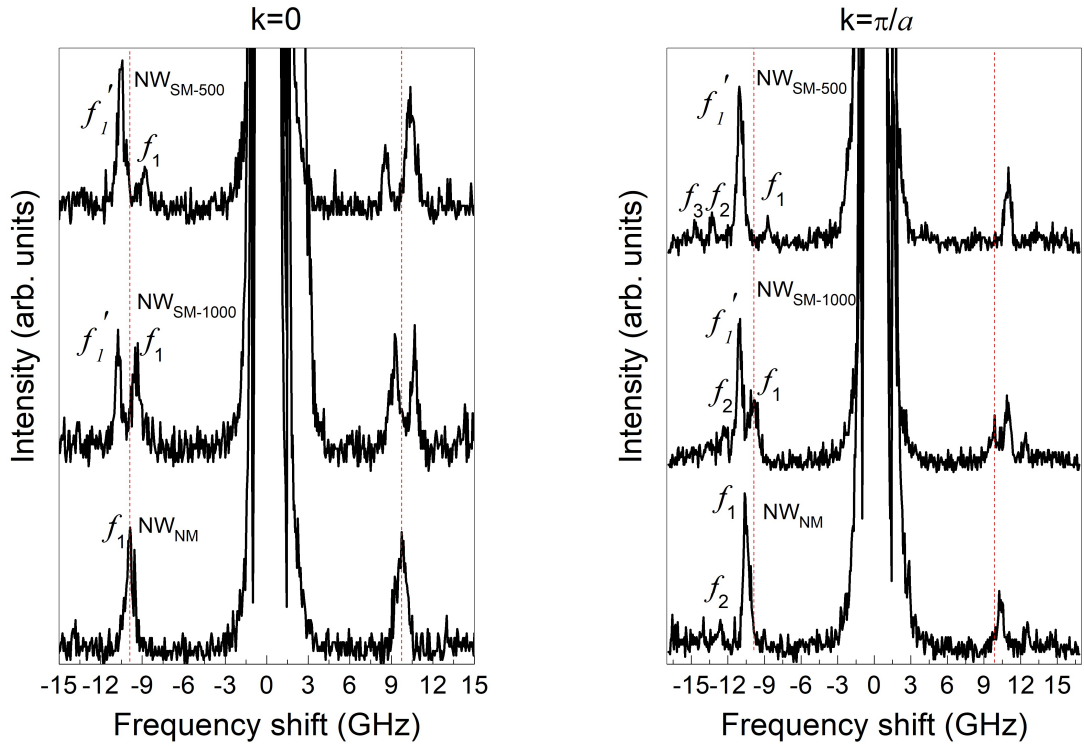

Fig. 2 BLS spectra measured at the center (Γ-point, i. e. $k=0$) and at the boundary (Y-point, i.e. $k=\pi/a$) of the first BZ for the three NWs investigated. The magnetic field H=500 Oe is applied along the NWs length (easy magnetization direction). Peaks are identified with the labels $f_1$, $f'_1$, $f_2$… put on the Stokes side of the measured spectra. The vertical dotted lines, centered on the frequency position (9.8 GHz) of the $f_1$ peak measured for $k=0$ in the $NW_{NM}$ array, are used as a guide to the eye.

The evolution of the frequency as a function of the *k*-vector for these peaks is plotted in Fig. 3 together with the DMM calculated dispersion. For the three NW arrays investigated, the dispersion is periodic with the appearance of BZs determined by the artificial periodicity of the NW array. Remarkably, the most intense modes, located in the lowest frequency part of the spectra are those exhibiting the largest amplitude of frequency oscillation. In all the cases, the experimental data are well reproduced by DMM calculations. Slight disagreement is primarily ascribed to the fact that we did not perform a real best-fit procedure of the experimental data to the calculated curves and that the same set of magnetic parameters is used to reproduce the dispersion of all the investigated arrays.

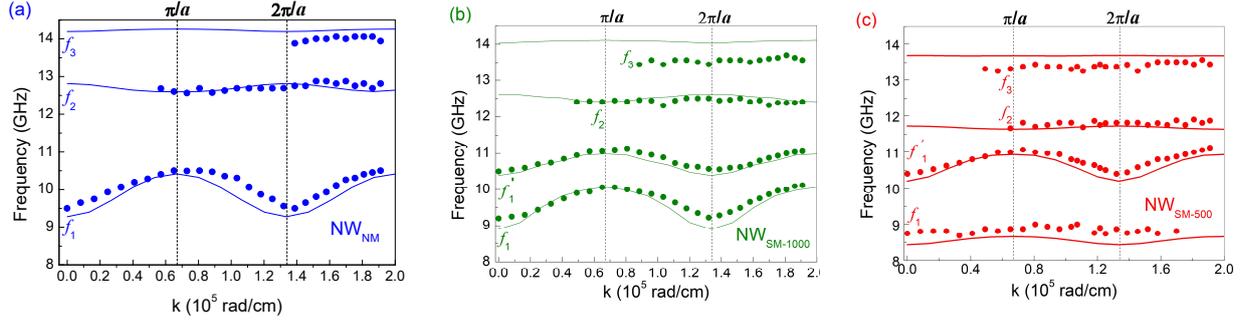

Fig. 3 Measured (points) and calculated (lines) frequency dispersion for homogeneous NWs width (a) and width modulated NWs with (b) 1000nm and (c) 500nm periodicity. The magnetic field H=500 Oe is applied along the length of NWs (easy magnetization direction). The vertical dashed lines mark the edge of the first BZ ($\pi/a$) and the center of the second BZ ($2\pi/a$).

For the $NW_{NM}$ only the lowest frequency mode ($f_1$) exhibits a sizeable dispersion with magnonic band width of about 1 GHz while the higher frequency modes ($f_2$ and $f_3$) are almost dispersionless. The principal consequence of the one side width-modulation in the $NW_{SM-1000}$ is the appearance of a second dispersive mode ($f_1'$) when compared to the results obtained for the homogeneous width NWs ($NW_{NM}$). This additional mode ($f_1'$) has higher frequency than $f_1$ and has comparable frequency oscillation amplitude with respect to mode $f_1$.

On the contrary for the $NW_{SM-500}$ array the lowest frequency mode ($f_1$) has negligible oscillation amplitude while the dispersion of the $f_1'$ mode is very similar to that observed in $NW_{SM-1000}$ both in the center frequency and amplitude of the magnonic band. In addition, the mode $f_2$ appears at a slight smaller frequency (about 0.5 GHz) than the corresponding mode observed in the $NW_{SM-1000}$ array. These results show that depending on the width-modulation periodicity ($p$) one can have a different number of propagative SWs with different frequency position and the width of the magnonic band.

Examination of the mode spatial profiles presented in Fig. 4 for the homogeneous NW width ($NW_{NM}$) reveal that they are the standard ones which are homogeneously distributed along the NWs length. At low frequency, the F mode has a quasi-uniform profile of the dynamic magnetization across the NW width (undulation in Fig. 4 is a numerical artifact) while the 1DE (Damon-Eshbach) and 2DE modes, exhibiting one and two nodal planes along the applied field direction, respectively, are observed at higher frequency. The F mode creates the largest dynamic dipolar field which efficiently couples a NW with its neighbors, resulting in a sizeable SW frequency dispersion for this mode.

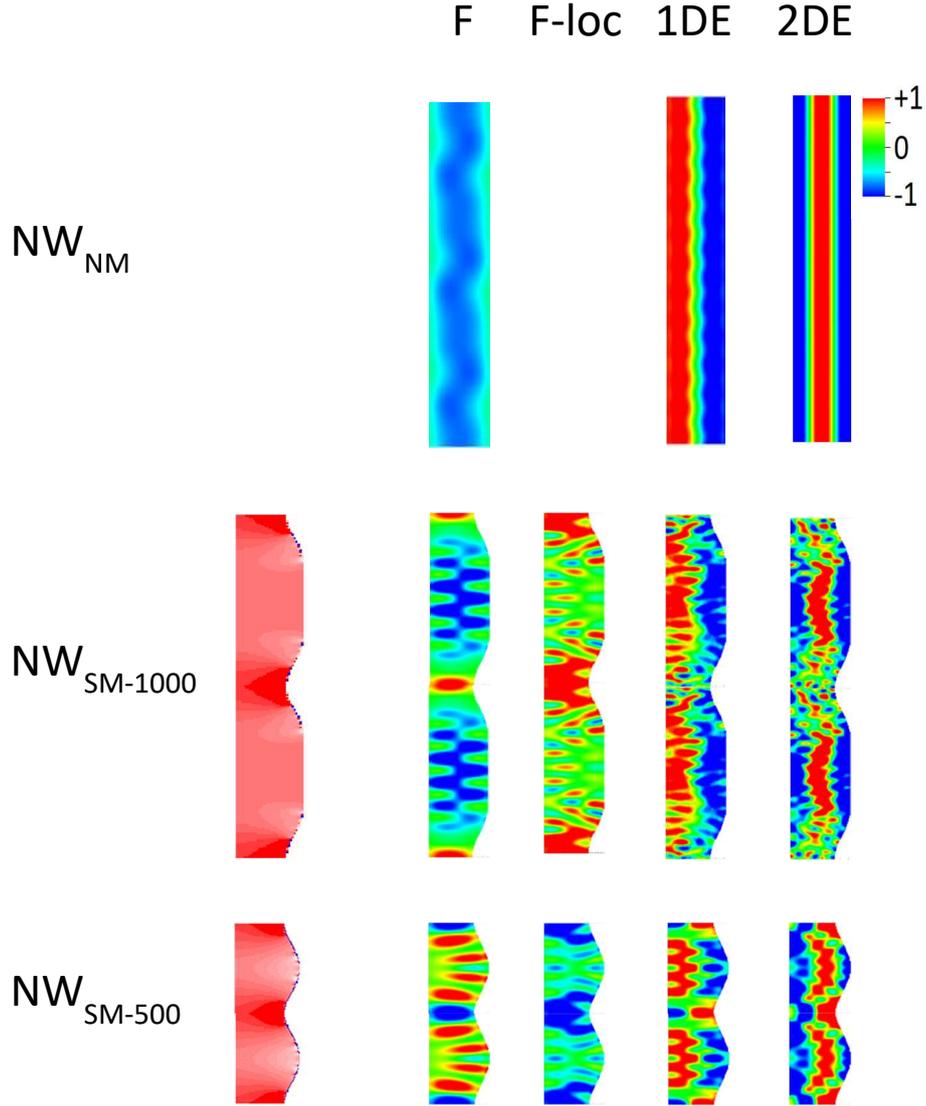

Fig. 4 Calculated spatial profiles (out-of-plane dynamic magnetization) of the principal modes detected by BLS for homogeneous with NWs ($NW_{NM}$) and single-side modulated NWs. The magnetic field H=500 Oe is applied along the length of NWs (easy magnetization direction). Red (white) regions correspond to maxima (minima) of the calculated effective magnetic field ($H_{eff}$) for width-modulated NWs. For the NW with homogeneous width $H_{eff}$ is uniform (not shown). The profiles of the modes are plotted in order of increasing frequency from left to right panels.

For the width-modulated NWs, an additional mode localized in the narrowest region of the NW, where the notch is present, is observed. For this reason, it is labelled as Fundamental-localized (F-loc) mode and can be considered as the quasi-uniform mode of the narrowest NW portion. The origin of this mode can be explained by inspection of the spatial profile, shown in Fig. 4, of the effective internal magnetic field $H_{eff}$ defined as the sum of the external, demagnetizing and exchange magnetic fields. For the $NW_{NM}$, $H_{eff}$ is uniform over the whole NW and this allows the existence of a mode with quasi-uniform (F) spatial profile. However, the introduction of the width-modulation forces the magnetization to curl, following the NW boundaries, and to misalign with respect H: in the widest NW portion the demagnetizing field is larger, and hence the $H_{eff}$ is lower (white/lighter areas in Fig. 4). As a consequence, $H_{eff}$ is characterized by alternating maxima and minima regions, representing "potential wells" for the SWs where the ideal fundamental mode of the homogeneous width NW, splits into two modes localized within regions with different $H_{eff}$ and consequently different frequencies.

We also notice that, for the $NW_{SM-1000}$, the F mode is much more uniform than that of the $NW_{SM-500}$ which exhibits pronounced amplitude oscillations and thus resulting in a smaller average dynamic magnetization which reduced the dipole stray field creates by this mode outside the NW itself. This corresponds to a reduction of the mode coupling through the array with a consequent reduction of the magnonic band amplitude which is clearly visible in Fig. 3.

In another paper, [31] two distinct resonances were observed by broadband ferromagnetic resonance in isolated (not interacting) width-modulated NWs in contrast with the single mode observed in the homogeneous width NWs. This study was limited to detect magnetization dynamics at the center of the first Brillouin zone (BZ) i.e., at a wave vector $k = 0$.

Finally, for all the NWs investigated, the 1DE ($f_3$) mode is detected with a frequency which decreases passing from the homogeneous NWs to the width-modulated NWs with increasing $p$ due to the non-uniform demagnetizing field while the 2DE ($f_4$) mode is only observed for the width-modulated NWs in the investigated frequency range (up to 14 GHz).

In this work, we have studied the SWs dispersion for arrays of closely spaced width-modulated nanowires and investigated its dependence on the width-modulation periodicity. For width-modulated nanowires with periodicity of $p$=1000 nm, two dispersive modes are observed in the lowest frequency range of the spectrum while for $p$=500 nm the lowest mode is almost dispersionless and the mode at higher frequency shows a significant amplitude of the magnonic band. These two modes have quasi-uniform spatial profiles and are localized into the widest and narrowest portions of the NWs. The reported results are important in view of tuning the band structure in one-dimensional magnonic crystal with multi-modal frequency transmission characteristics.

A.O.A. was supported by the National Research Foundation, Prime Minister's Office, Singapore under its Competitive Research Programme (CRP Award No. NRFCRP 10-2012-03).